\documentclass[aps,prd,onecolumn,groupedaddress,showpacs,nofootinbib,amssymb, preprint]{revtex4}
\usepackage[dvips]{graphicx}
\usepackage{amssymb}
\usepackage{amsmath}
\usepackage{graphicx,,color}
\usepackage{amsfonts}
\usepackage{bm}
\usepackage{cancel}
\usepackage{comment}
\usepackage{floatflt}
\usepackage{slashed}

\newcommand\be{\begin{align}}
\newcommand\ee{\end{align}}

\newcommand\e{\mathrm{e}}

\allowdisplaybreaks[4]
\begin{document}
\title{Phantom Crossing and Oscillating Dark Energy with $F(R)$ Gravity}
\author{S.~Nojiri,$^{1,2}$}
\email{nojiri@nagoya-u.jp}
\author{S.D.~Odintsov,$^{3,4}$}
\email{odintsov@ice.cat}
\author{V.K.~Oikonomou,$^{5}$}
\email{v.k.oikonomou1979@gmail.com;voikonomou@gapps.auth.gr}
\affiliation{$^{1)}$ Theory Center, High Energy Accelerator Research Organization (KEK), \\
Oho 1-1, Tsukuba, Ibaraki 305-0801, Japan \\
$^{2)}$ Kobayashi-Maskawa Institute for the Origin of Particles
and the Universe, Nagoya University, Nagoya 464-8602, Japan \\
$^{3)}$ ICREA, Passeig Luis Companys, 23, 08010 Barcelona, Spain\\
$^{4)}$ Institute of Space Sciences (IEEC-CSIC) C. Can Magrans
s/n, 08193 Barcelona, Spain\\
$^{5)}$ Department of Physics, Aristotle University of
Thessaloniki, Thessaloniki 54124, Greece}

\begin{abstract}
In this work, we shall consider how a dynamical oscillating and phantom crossing dark energy era can be realized in the context of $F(R)$ gravity. 
We approach the topic from a theoretical standpoint considering all the conditions that may lead to a consistent phantom crossing behavior and separately how the $F(R)$ gravity context may realize oscillating dark energy era. 
Apart from our qualitative considerations, we study in a quantitative way two $F(R)$ gravity dark energy models which are viable cosmologically and also exhibit simultaneously phantom crossing behavior and also oscillating dark energy. 
We consider these models by solving numerically the field equations using appropriate statefinder parameters engineered for dark energy studies. 
As we show, $F(R)$ provides a natural extension of Einstein's general relativity which can naturally realize a transition from a phantom era to a quintessential era, a feature supported by recent observational data, without resorting to phantom scalar fields to realize the phantom evolution.
\end{abstract}

\maketitle

\section{Introduction}

The observation of the late-time acceleration of the Universe,
known as dark energy (DE) era is one of the most unexpected
features of our Universe. In the context of simple general
relativity, the DE era can be generated by quintessence scalar
fields, however the latest observations of the Dark Energy
Spectroscopic Instrument (DESI) \cite{DESI:2024mwx,DESI:2025zgx}
indicate that the DE was in the phantom regime in the near past.
This feature can be described in the context of general relativity
only by using phantom scalar fields, which is a rather
unattractive feature. Also and more importantly, the DESI data
indicate a transition from a phantom to a quintessential regime
for DE. This is questionable if it can be realized in the context
of general relativity. There are many ways to produce results
compatible with the DESI data, for example
\cite{OdintsovSGS:2024,Dai:2020rfo,He:2020zns,Nakai:2020oit,DiValentino:2020naf,Agrawal:2019dlm,Ye:2020btb,Vagnozzi:2021tjv,
Desmond:2019ygn,Hogas:2023pjz,OColgain:2018czj,Vagnozzi:2019ezj,
Krishnan:2020obg,Colgain:2019joh,Vagnozzi:2021gjh,Lee:2022cyh,Krishnan:2021dyb,Ye:2021iwa,Ye:2022afu,Verde:2019ivm,Menci:2024rbq,Adil:2023ara,
Reeves:2022aoi,Ferlito:2022mok,Vagnozzi:2021quy,DiValentino:2020evt,Sabogal:2024yha,Giare:2024smz,DiValentino:2025sru,Odintsov:2025kyw,vanderWesthuizen:2025iam,Paliathanasis:2025dcr,Odintsov:2025jfq,Kessler:2025kju}
and among these frameworks, $F(R)$ gravity
\cite{reviews1,reviews2,reviews3,reviews4,Nojiri:2003ft,Capozziello:2005ku,Hwang:2001pu,Song:2006ej,Faulkner:2006ub,Olmo:2006eh,Sawicki:2007tf,Faraoni:2007yn,Carloni:2007yv,
Nojiri:2007as,Deruelle:2007pt,Appleby:2008tv,Dunsby:2010wg,Oikonomou:2025qub}
enjoys an elevated role. There are recent studies which discuss
the ability of $F(R)$ gravity to be compatible with the DESI data
\cite{Odintsov:2025jfq}. Another more fascinating behavior of DE
based on the DESI data was demonstrated in Ref.
\cite{Kessler:2025kju}, which indicates that the DE might not be
only dynamical and evolving from phantom to quintessence, but it
also shows strong indications of an oscillating equation of state
(EoS). In this work we aim to investigate whether a phantom
involving and oscillating DE may be realized by $F(R)$ gravity in
general. We approach the problem theoretically, finding the
conditions which may allow such a phenomenological behavior to be
produced and also we present several phenomenological models from
a theoretical point of view. In addition, we study numerically the
DE era of some explicit $F(R)$ gravity models which can realize a
viable Planck compatible \cite{Planck:2018vyg} and in some cases
DESI compatible late-time behavior. We also demonstrate that these
models produce an oscillating DE era and also allow for phantom to
quintessence transitions.

This paper is organized as follows: In section II we discuss the
inverse-phantom prossing in $F(R)$ gravity from a theoretical
viewpoint. The analysis is continued in section III where we
discuss the apparent phantom crossing. In section IV, the
theoretical study of several models that exhibit phantom crossing
is considered. The oscillating Hubble rate is considered in
section V, while in section VI the effects of a scalaron as dark
matter with curvature-dependent mass on the DE era are considered.
In section VII, we discuss the features of a realistic model while
in section VIII we numerically study two explicit examples of
oscillating $F(R)$ gravity DE with phantom crossing. Finally, the
conclusions follow at the end of the article.

\section{Inverse-Phantom Crossing in $F(R)$ gravity}\label{SecII}

Consider the vacuum $F(R)$ gravity action,
\begin{align}
\label{actiondevac} \centering \mathcal{S}=\frac{1}{2\kappa^2}\int
d^4x\sqrt{-g}F(R)\, ,
\end{align}
and a flat Friedmann-Lema\^{i}tre-Robertson-Walker (FLRW) metric with the
following line element,
\begin{align}
 \centering\label{frw}
 d s^2 = - d t^2 + a(t) \sum_{i = 1}^3 d x_i^2\, .
\end{align}
Let $w$ be the DE equation of state parameter, that is, the ratio
of the pressure of the DE fluid with respect to the energy density
of the DE fluid. If the recent DESI observations
\cite{DESI:2024mwx, DESI:2025zgx} are correct, there might be a
transition from $w<-1$ to $w>-1$ in the DE EoS.

In this section, we discuss if the initial condition determining
if the standard phantom crossing (from $w>-1$ to $w<-1$) or the
inverse phantom crossing (from $w<-1$ to $w>-1$) can be realized
theoretically in the context of $F(R)$ gravity. Therefore, we show
that the phantom crossing  (or inverse phantom crossing) in the
$F(R)$ sector (which is described by the effective pressure and
the effective energy density) can occur for almost any $F(R)$
gravity by adjusting the initial conditions.

The conservation law for a FLRW spacetime is,
\begin{align}
\label{c}
0 = \dot\rho + 3 H \left( \rho + p \right)\, ,
\end{align}
which indicates that when the phantom crossing occurs, since we
have $\rho=-p$, then $\dot\rho$ must vanish. Due to the fact that
in the phantom Universe, $\rho<-p$, $\dot \rho$ and in the
non-phantom Universe, $\rho>p$, there is a crossing from the
non-phantom Universe to the phantom one if $\ddot\rho>0$ and there
is an inverse crossing if $\ddot\rho<0$.

Let the $F(R)$ gravity be of the form,
\begin{align}
\label{FRf}
F(R) = R + f(R) \, .
\end{align}
The effective energy density $\rho_{f(R)}$ and the
effective pressure $p_{f(R)}$ coming from the $f(R)$ part
are given by,
\begin{align}
\label{Cr4}
\rho_{f(R)} =&\,
\frac{1}{\kappa^2}\left(-\frac{1}{2}f(R) + 3\left(H^2
+ \dot H\right) f'(R) - 18 \left(4H^2 \dot H
+ H \ddot H\right)f''(R)\right) 
\, ,\\
\label{Cr4bb}
p_{f(R)} =&\,
\frac{1}{\kappa^2}\left(\frac{1}{2}f(R) - \left(3H^2
+ \dot H \right)f'(R) + 6 \left(8H^2 \dot H + 4{\dot H}^2 + 6 H \ddot H
+ \dddot H \right)f''(R) \right. \nonumber \\
&\, \left. + 36\left(4H\dot H + \ddot H\right)^2f'''(R)
\right) 
\, .
\end{align}
We consider the phantom crossing, where $\dot \rho_{f(R)} =
\rho_{f(R)} + p_{f(R)} =0$, in the $f(R)$ sector by
using (\ref{Cr4}) and (\ref{Cr4bb}). The Friedmann equation is
given by,
\begin{align}
\label{1F} \frac{3}{\kappa^2} H^2 = \rho_{f(R)} +
\rho_\mathrm{matter} \, ,
\end{align}
where we also considered the contribution of perfect matter
fluids. We now assume that the matter energy density
$\rho_\mathrm{matter}$ is given by the dust, that is, the baryonic
matter and the cold dark matter. Then we may assume,
\begin{align}
\label{rhom}
\rho_\mathrm{matter} = \rho_\mathrm{matter(0)} a^{-3} = \rho_\mathrm{matter(0)} \e^{-3N} \, .
\end{align}
Here $\rho_\mathrm{matter(0)}$ is a constant and $a$ is the scale
factor, which is given in terms of the $e$-foldings number $N$ as
$a=\e^N$.

When the phantom crossing in the $f(R)$ sector happens at the
cosmic time instance $t=t_\mathrm{ph}$, we can assume that $N$ can
be expanded as follows,
\begin{align}
\label{phN}
N=\sum_{n=0} \frac{H_n}{\left( n + 1 \right)!} \left( t - t_\mathrm{ph} \right)^{n+1}\, ,
\end{align}
which gives the following Hubble rate $H$,
\begin{align}
\label{phH}
H=\sum_{n=0} \frac{H_n}{n !} \left( t - t_\mathrm{ph} \right)^n\, .
\end{align}
Then we find that the scalar curvature $R$ behaves as,
\begin{align}
\label{phR}
R =&\, 12 H^2 + 6\dot H
= 12 {H_0}^2 + 6 H_1 + \left( 24 H_0 H_1 + 6 H_2 \right) \left( t - t_\mathrm{ph} \right) \nonumber \\
&\, + \left( 12 H_0 H_2 + 12 {H_1}^2 + 3 H_3 \right) \left( t - t_\mathrm{ph} \right)^2
+ \mathcal{O} \left( \left( t - t_\mathrm{ph} \right)^3 \right)\, .
\end{align}
We also assume that,
\begin{align}
\label{rhofR}
\rho_{f(R)} = \rho_{{f(R)}\, (0)} + \rho_{{f(R)}\, (2)} \left( t - t_\mathrm{ph} \right)^2
+ \mathcal{O} \left( \left( t - t_\mathrm{ph} \right)^3 \right)\, .
\end{align}
Then if $\rho_{{f(R)}\, (2)}>0$, a crossing from the non-phantom $\rho_{f(R)} > p_{f(R)}$
to the phantom $\rho_{f(R)} < p_{f(R)}$ occurs and if $\rho_{{f(R)}\, (2)}<0$,
an inverse crossing occurs.

The first Friedmann equation (\ref{1F}) indicates that,
\begin{align}
\label{1F0}
\frac{3}{\kappa^2} {H_0}^2 =&\, \rho_{{f(R)}\, (0)} + \rho_\mathrm{matter(0)} \, , \\
\label{1F1}
\frac{6}{\kappa^2} H_0 H_1 =&\, - 3 \rho_\mathrm{matter(0)} H_0 \, , \\
\label{1F2}
\frac{3}{\kappa^2} \left( H_0 H_2 + 2 {H_1}^2 \right) =&\, \rho_{{f(R)}\, (2)} + \rho_\mathrm{matter(0)} \left( \frac{9}{2} {H_0}^2 - \frac{3}{2} H_1 \right) \, .
\end{align}
On the other hand, Eq.~(\ref{Cr4}) tells,
\begin{align}
\label{Cr4t1}
\rho_{f(R)\, (0)} =&\,
\frac{1}{\kappa^2}\Bigl(-\frac{1}{2}f\left( 12 {H_0}^2 + 6 H_1 \right)
+ 3\left( {H_0}^2 + H_1 \right) f'\left( 12 {H_0}^2 + 6 H_1 \right) \nonumber \\
&\, - 18 \left(4{H_0}^2 H_1 + H_0 H_2 \right)f''\left( 12 {H_0}^2 + 6 H_1 \right) \Bigr) 
\, ,\\
\label{rhofR1}
0 
=&\,
\frac{1}{\kappa^2}\left\{
 - 6 H_0 H_1 f'\left( 12 {H_0}^2 + 6 H_1 \right) \right. \nonumber \\
&\, + 18 \left( 4 {H_0}^3 H_1 - 3 {H_0}^2 H_2 - 4 H_0 {H_1}^2 - H_0 H_3 \right) f''\left( 12 {H_0}^2 + 6 H_1 \right) \nonumber \\
&\, \left.
 - 108 H_0 \left(4 H_0 H_1 + H_2 \right)^2 f'''\left( 12 {H_0}^2 + 6 H_1 \right) \right\}
\, ,\\
\label{rhofR2}
\rho_{f(R)\, (2)}
=&\,
\frac{1}{\kappa^2}\left\{ - 9 H_0 H_2 f'\left( 12 {H_0}^2 + 6 H_1 \right) \right. \nonumber \\
&\, + \left( - 36 {H_0}^2 H_2 + 108 H_0 H_1 H_2
 - 144 {H_1}^3 + 9 {H_2}~2 \right) f''\left( 12 {H_0}^2 + 6 H_1 \right) \nonumber \\
&\, + \left( 24 H_0 H_1 + 6 H_2 \right) \left( 36 {H_0}^3 H_1 - 63 {H_0}^2 H_2 - 108 H_0 {H_1}^2
 - 18 H_0 H_3 - 9 H_1 H_2 \right) \nonumber \\
&\, \times f'''\left( 12 {H_0}^2 + 6 H_1 \right)
\nonumber \\
&\, \left.
 - 324 H_0 \left( 4 H_0 H_1 + H_2 \right)^3 f''''\left( 12 {H_0}^2 + 6 H_1 \right) \right\} 
\, .
\end{align}
In the following, we consider the solution of the above equations
and their physical meanings. Eq.~(\ref{1F1}) indicates that,
\begin{align}
\label{H1} H_1 = - \frac{\kappa^2}{2} \rho_\mathrm{matter(0)} \, ,
\end{align}
therefore, $H_1$ is always negative. This expression can also be
obtained by combining the first and second Friedmann equations,
\begin{align}
\label{H1B}
 - \frac{2}{\kappa^2} \dot H = \rho_{f(R)} + p_{f(R)} + \rho_\mathrm{matter} + p_\mathrm{matter} \, ,
\end{align}
and noting $\rho_{f(R)} + p_{f(R)} =0$ when the phantom crossing occurs for $f(R)$ sector and
$p_\mathrm{matter} = 0$ for cold matter and baryonic matter.

By combining (\ref{1F0}), (\ref{Cr4t1}), and (\ref{H1}), we find
the following expression for $H_2$,
\begin{align}
\label{H2}
H_2 =&\, - 4 H_0 H_1 - \frac{f\left( 12 {H_0}^2 + 6 H_1 \right)}{36 H_0 f''\left( 12 {H_0}^2 + 6 H_1 \right)}
+ \frac{\left( {H_0}^2 + H_1 \right) f'\left( 12 {H_0}^2 + 6 H_1 \right)}{6H_0 f''\left( 12 {H_0}^2 + 6 H_1 \right)} \nonumber \\
&\, - \frac{3{H_0}^2 + H_1}{18 H_0 f''\left( 12 {H_0}^2 + 6 H_1 \right)} \, .
\end{align}
Eq.~(\ref{1F2}) with (\ref{H1}) and (\ref{H2}) indicate that,
\begin{align}
\label{rho2}
\rho_{{f(R)}\, (2)} =&\, \frac{3}{\kappa^2} \left( H_0 H_2 + 2 {H_1}^2 \right)
+ \frac{2}{\kappa^2} H_1 \left( \frac{9}{2} {H_0}^2 - \frac{3}{2} H_1 \right) \nonumber \\
=&\, \frac{3}{\kappa^2} \left( - {H_0}^2 H_1 + {H_1}^2 \right)
 - \frac{f\left( 12 {H_0}^2 + 6 H_1 \right)}{12 \kappa^2 f''\left( 12 {H_0}^2 + 6 H_1 \right)} \nonumber \\
&\, + \frac{\left( {H_0}^2 + H_1 \right) f'\left( 12 {H_0}^2 + 6 H_1 \right)}{2 \kappa^2 f''\left( 12 {H_0}^2 + 6 H_1 \right)}
 - \frac{3{H_0}^2 + H_1}{6 \kappa^2 f''\left( 12 {H_0}^2 + 6 H_1 \right)} \, .
\end{align}
Therefore, it depends on the detailed form of $f(R)$ if a crossing
from the non-phantom to the phantom occurs, that is,
$\rho_{{f(R)}\, (2)}>0$, or an inverse crossing, that is,
$\rho_{{f(R)}\, (2)}<0$, occurs. This could indicate that
the phantom crossing or the inverse phantom crossing occurs in a
wide class of $F(R)$ gravity models, depending on the initial
condition determining $H_0$. Furthermore, by combining
(\ref{rhofR2}) with (\ref{rho2}), we can find the expression of
$H_3$ although we do not write it explicitly because we do not use
the expression of $H_3$ later.

As an example, we may consider $R^2$ gravity,
\begin{align}
\label{R2} f(R)=\alpha R^2 \, ,
\end{align}
where $\alpha$ is a constant. As is well-known, the $F(R)$ gravity
includes a scalar mode. In the model (\ref{R2}), the square of the
mass is given by $\frac{1}{\alpha}$. Therefore, in order to avoid
the tachyon, we require $\alpha$ to be positive.

For the model (\ref{R2}), Eq.~(\ref{rho2}) has the following form,
\begin{align}
\label{rho2B0}
\rho_{{f(R)}\, (2)}
=&\, \frac{1}{2\kappa^2} \left( 9 {H_1}^2 - \frac{3{H_0}^2 + H_1}{6 \alpha} \right) \, .
\end{align}
For the standard phantom crossing to occur, the condition is given
by,
\begin{align}
\label{sphcndtn} {H_0}^2 < 18 \alpha {H_1}^2 - \frac{1}{3} H_1 =
\frac{9 \alpha \kappa^4}{2} {\rho_\mathrm{matter(0)}}^2 +
\frac{\kappa^2}{6} \rho_\mathrm{matter(0)} \, ,
\end{align}
were we used Eq. (\ref{H1}). On the other hand, the condition for
the inverse phantom crossing is given by,
\begin{align}
\label{iphcndtn}
{H_0}^2 > 18 \alpha {H_1}^2 - \frac{1}{3} H_1 = \frac{9 \alpha \kappa^4}{2} {\rho_\mathrm{matter(0)}}^2
+ \frac{\kappa^2}{6} \rho_\mathrm{matter(0)} \, .
\end{align}
Then the crossing depends on the initial condition to determine $H_0$ and $\rho_\mathrm{matter(0)}$.

\section{Apparent Phantom Crossing}\label{SecIII}

There is an idea that instead of considering the transition from
$w<-1$ to $w>-1$ of DE, the modification of the dark matter sector
may solve the problem in the DESI observations
\cite{Khoury:2025txd}. If dark matter density decreases more
slowly than  $a^{-3}$, as usually predicted from energy
conservation, it looks like there could be a phantom crossing in
the DE sector because we are considering the total energy density,
and we assume the usual  $a^{-3}$ behavior of the dark matter for
the analysis of the DESI observations.

If the dark matter is non-relativistic, as in the cold dark
matter, the energy density of the dark matter is given by the
product $\rho_\mathrm{DM} = m_\mathrm{DM} n_\mathrm{DM}$ of the
mass of the dark matter particle, $m_\mathrm{DM}$, and the number
density of the dark matter particle, $n_\mathrm{DM}$. If there is
neither creation nor annihilation of the dark matter particles,
the number density $n_\mathrm{DM}$ is proportional to the inverse
of the volume in the expanding Universe, $n_\mathrm{DM}\propto
V^{-1} \propto a^{-3}$. If the mass $m_\mathrm{DM}$ is time
dependent $m_\mathrm{DM} (t)$ or $m_\mathrm{DM}$ depends on the
scalar curvature $R$, the behavior of the dark matter energy
density $\rho_\mathrm{DM}$ is different from $a^{-3}$. If the mass
depends on the scalar curvature, the energy density also includes
the derivative terms because the curvature includes the
derivative, which makes the situation somewhat complicated.

In order to make the physical problem at hand well-defined, we
assume that the dark matter is a Dirac spinor, whose action is
given by,
\begin{align}
\label{DMsp}
S_\mathrm{DM}= \int d^4 x \sqrt{-g} \left\{ i \bar\psi \slashed{D} \psi - m(R) \bar\psi \psi \right\} \, .
\end{align}
Here $\slashed{D} \equiv \gamma^\mu D_\mu$ and $D_\mu$ is the
covariant derivative including spin connection. For the action
(\ref{DMsp}), the Hamiltonian density $\mathcal{H}$ is given by,
\begin{align}
\label{spH}
\mathcal{H} = \bar\psi \left( - i \gamma^i D_i + m(R) \right) \psi\, ,
\end{align}
in the canonical formulation. The energy density is the quantum or
statistical average of the Hamiltonian density, as follows,
\begin{align}
\label{sprho}
\rho_\mathrm{DM} = \left< \mathcal{H} \right> = -i \left< \bar\psi \gamma^i \partial_i \psi \right> + m(R) \left< \bar\psi \psi \right> \, .
\end{align}
The first term corresponds to the kinetic energy and the second
term to the rest mass. As long as we consider non-relativistic
particles, the first term can be neglected by comparing the second
term. We should also note that $\left< \bar\psi \psi \right>$ is
nothing but the number density of dark matter $n_\mathrm{DM}$.
Then we obtain,
\begin{align}
\label{sprho2}
\rho_\mathrm{DM} = m(R) n_\mathrm{DM} \, ,
\end{align}
as expected. Hence, if we choose $m(R)$ to be a decreasing
function of $R$, at late times where $R$ decreases, $m(R)$ becomes
larger and $\rho_\mathrm{DM}$ decreases slower than $a^{-3}$.

The $R$-dependence of the mass, makes the situation complicated
when we consider the variation with respect to the metric or the
vierbein field. The variation of the action (\ref{DMsp}) with
respect to the metric or the vierbein field gives the following
energy-momentum tensor of the Dirac field,
\begin{align}
\label{EMDiracR}
T_{\mu\nu}=&\, \frac{i}{2} \left( \bar\psi \gamma_\mu D_\nu \psi - \left( D_\nu \bar\psi\right) \gamma_\mu \psi \right)
 - g_{\mu\nu} \left\{ i \bar\psi \slashed{D} \psi - m(R) \bar\psi \psi \right\} \nonumber \\
&\, + 2\left( - R_{\mu\nu} - g_{\mu\nu} \Box + \nabla_\mu \nabla_\nu \right) \left( m'(R) \bar\psi \psi \right) \, .
\end{align}
On the other hand, the Dirac equations given by the action
(\ref{DMsp}) with respect to the Dirac spinors, $\psi$ and
$\bar\psi$ have the following forms,
\begin{align}
\label{Diraceqs}
0 = i \slashed{D} \psi - m(R) \psi \, , \quad
0 = - i \left( D_\mu \bar\psi\right) \gamma^\mu - m(R) \bar\psi \, .
\end{align}
We assume that $\left< \bar\psi \gamma_i D_j \psi \right>$ and
$\left< \left( D_i \bar\psi\right) \gamma_j \psi \right>$ are
negligible for the non-relativistic dark matter. Then we find the
energy density $\tilde \rho_\mathrm{DM}$ and the pressure $\tilde
p_\mathrm{DM}$ as follows,
\begin{align}
\label{rhopDM}
\tilde\rho_\mathrm{DM} =&\, m(R) \left< \bar\psi \psi \right>
+ \left\{ 3\left(\dot H + H^2\right) + 3H \frac{d}{dt} \right\} \left( m'(R) \left< \bar\psi \psi \right> \right) \nonumber \\
=&\, m(R) n_\mathrm{DM}+ \left\{ 3\left(\dot H + H^2\right) + 3H \frac{d}{dt} \right\} \left( m'(R) n_\mathrm{DM}\right) \, , \nonumber \\
\tilde p_\mathrm{DM} =&\, \left\{ - \left(\dot H + 3 H^2\right) + \frac{d^2}{dt^2} + 2 H \frac{d}{dt} \right\} \left( m'(R) \left< \bar\psi \psi \right> \right) \nonumber \\
=&\, \left\{ - \left(\dot H + 3 H^2\right) + \frac{d^2}{dt^2} + 2 H \frac{d}{dt} \right\} \left( m'(R) n_\mathrm{DM} \right) \, .
\end{align}
The expression of $\tilde\rho_\mathrm{DM}$ in (\ref{rhopDM}) is different from $\rho_\mathrm{DM}$ in (\ref{sprho2}) by the second term.

By combining $\tilde\rho_\mathrm{DM}$ and $\tilde p_\mathrm{DM}$
with $\rho_{f(R)}$ and the effective pressure
$p_{f(R)}$ coming from $f(R)$ term in (\ref{Cr4}) and
(\ref{Cr4bb}), we find the total effective energy density
$\rho_\mathrm{total}\equiv \tilde\rho_\mathrm{DM} +
\rho_{f(R)}$ and the total pressure $p_\mathrm{total}\equiv
\tilde p_\mathrm{DM} + p_{f(R)}$, which satisfy the
standard conservation law, $0=\dot \rho_\mathrm{total} + 3 H
\left( \rho_\mathrm{total} + p_\mathrm{total} \right)$. We may
subtract the part which behaves as $a^{-3}$ from the effective
energy density $\rho_\mathrm{eff\, DE}$ of the DE,
\begin{align}
\label{effDE}
\rho_\mathrm{eff\, DE} \equiv \rho_\mathrm{total} - \rho_0 a^{-3}\, .
\end{align}
Here $\rho_0$ is a constant corresponding to the energy density of
the dark matter in the present Universe if we choose $a=1$ in the
present Universe. We may also identify the effective pressure
$p_\mathrm{eff\, DE}$ of DE with the total pressure
$p_\mathrm{total}$, $p_\mathrm{eff\, DE}=p_\mathrm{total}$. Then
the effective energy and the effective pressure satisfy the
conservation law, $0=\dot \rho_\mathrm{eff\, DE} + 3 H \left(
\rho_\mathrm{eff\, DE} + p_\mathrm{eff\, DE}\right)$. Then even if
the effective energy density $\rho_{f(R)}$ and the pressure
$p_{f(R)}$ of the $f(R)$ sector do not show the phantom
crossing, the effective energy density $\rho_\mathrm{eff\, DE}$
and the effective pressure $p_\mathrm{eff\, DE}$ of DE may appear
to cause the phantom crossing.

\section{Several Models that Exhibit Phantom Crossing: A Theoretical Account}\label{SecIV}

We now consider some specific examples that exhibit DE phantom
crossing. We may consider the model \cite{Odintsov:2024woi},
\begin{align}
\label{exp} F(R)=R - 2\Lambda \left( 1 - \e^{- \beta \left(
\frac{R}{2\Lambda}\right)^\alpha} \right)\, ,
\end{align}
where $\Lambda$ is a constant with dimension $[m]^2$ and $\beta$
is a dimensionless constant. When $R$ is large, $\beta
\left(\frac{R}{2\Lambda}\right)^\alpha \gg 1$ if $\alpha>0$, and
$\Lambda$ plays the role of the cosmological constant. This model
is also proposed to solve the inverse phantom crossing. We choose
$\frac{\Lambda}{\beta^\frac{1}{\alpha}}$ small enough compared
with the present value of the scalar curvature. This makes
$\Lambda$ to realize a DE era in the present Universe. In order to
also describe the inflation in the early Universe, we may add the
term $\gamma R^2$ in the $F(R)$ of (\ref{exp}) with a constant
$\gamma$.

Because $f(R)=- 2\Lambda \left( 1 - \e^{- \beta \left(
\frac{R}{2\Lambda}\right)^\alpha} \right)$, we find,
\begin{align}
\label{fdfdd}
f'(R) =&\, - \alpha\beta \left( \frac{R}{2\Lambda}\right)^{\alpha -1} \e^{- \beta \left( \frac{R}{2\Lambda}\right)^\alpha} \, , \nonumber \\
f'(R) =&\, - \frac{\alpha\beta}{2\Lambda} \left\{ \left(\alpha -1 \right) \left( \frac{R}{2\Lambda}\right)^{\alpha -2}
 - \alpha\beta \left( \frac{R}{2\Lambda}\right)^{2\alpha - 2} \right\} \e^{- \beta \left( \frac{R}{2\Lambda}\right)^\alpha} \, .
\end{align}
Then $\rho_{{f(R)}\, (2)}$ in (\ref{rho2}) has the following form,
\begin{align}
\label{rho2B}
\rho_{{f(R)}\, (2)}
=&\, \frac{3}{\kappa^2} \left( - {H_0}^2 H_1 + {H_1}^2 \right)
 - \frac{ 2\Lambda \left( 1 - \e^{- \beta \left( \frac{12 {H_0}^2 + 6 H_1 }{2\Lambda}\right)^\alpha} \right)
\e^{ \beta \left( \frac{12 {H_0}^2 + 6 H_1 }{2\Lambda}\right)^\alpha}}{12 \kappa^2
\frac{\alpha\beta}{2\Lambda} \left\{ \left(\alpha -1 \right) \left( \frac{12 {H_0}^2 + 6 H_1}{2\Lambda}\right)^{\alpha -2}
 - \alpha\beta \left( \frac{12 {H_0}^2 + 6 H_1}{2\Lambda}\right)^{2\alpha - 2} \right\} } \nonumber \\
&\, + \frac{\left( {H_0}^2 + H_1 \right) \alpha\beta \left( \frac{12 {H_0}^2 + 6 H_1}{2\Lambda}\right)^{\alpha -1} }{2 \kappa^2
\frac{\alpha\beta}{2\Lambda} \left\{ \left(\alpha -1 \right) \left( \frac{12 {H_0}^2 + 6 H_1}{2\Lambda}\right)^{\alpha -2}
 - \alpha\beta \left( \frac{12 {H_0}^2 + 6 H_1}{2\Lambda}\right)^{2\alpha - 2} \right\}} \nonumber \\
&\, + \frac{\left( 3{H_0}^2 + H_1 \right)\e^{ \beta \left( \frac{12 {H_0}^2 + 6 H_1 }{2\Lambda}\right)^\alpha}}{6 \kappa^2
\frac{\alpha\beta}{2\Lambda} \left\{ \left(\alpha -1 \right) \left( \frac{12 {H_0}^2 + 6 H_1}{2\Lambda}\right)^{\alpha -2}
 - \alpha\beta \left( \frac{12 {H_0}^2 + 6 H_1}{2\Lambda}\right)^{2\alpha - 2} \right\}} \nonumber \\
=&\, \frac{3}{\kappa^2} \left( - {H_0}^2 H_1 + {H_1}^2 \right)
 - \frac{ \Lambda^2 \left( 1 - \e^{- \beta \left( \frac{12 {H_0}^2 + 6 H_1 }{2\Lambda}\right)^\alpha} \right)
\e^{ \beta \left( \frac{12 {H_0}^2 + 6 H_1 }{2\Lambda}\right)^\alpha}}{3 \alpha\beta\kappa^2
\left\{ \left(\alpha -1 \right) \left( \frac{12 {H_0}^2 + 6 H_1}{2\Lambda}\right)^{\alpha -2}
 - \alpha\beta \left( \frac{12 {H_0}^2 + 6 H_1}{2\Lambda}\right)^{2\alpha - 2} \right\} } \nonumber \\
&\, + \frac{\Lambda \left( {H_0}^2 + H_1 \right) \left( \frac{12 {H_0}^2 + 6 H_1}{2\Lambda}\right)^{\alpha -1} }{\kappa^2
\left\{ \left(\alpha -1 \right) \left( \frac{12 {H_0}^2 + 6 H_1}{2\Lambda}\right)^{\alpha -2}
 - \alpha\beta \left( \frac{12 {H_0}^2 + 6 H_1}{2\Lambda}\right)^{2\alpha - 2} \right\}} \nonumber \\
&\, + \frac{\Lambda \left( 3{H_0}^2 + H_1 \right)\e^{ \beta \left( \frac{12 {H_0}^2 + 6 H_1 }{2\Lambda}\right)^\alpha}}{3 \alpha\beta\kappa^2
\left\{ \left(\alpha -1 \right) \left( \frac{12 {H_0}^2 + 6 H_1}{2\Lambda}\right)^{\alpha -2}
 - \alpha\beta \left( \frac{12 {H_0}^2 + 6 H_1}{2\Lambda}\right)^{2\alpha - 2} \right\}} \, .
\end{align}
Then, depending on the choice of the values of $H_0$ and $H_1$ or
dark matter energy density, the phantom crossing or the inverse
phantom crossing occurs. Especially, we choose the simple case
$\alpha=1$. Then Eq.~(\ref{rho2B}) reduces to,
\begin{align}
\label{rho2C}
\rho_{{f(R)}\, (2)}
=&\, \frac{3}{\kappa^2} \left( - {H_0}^2 H_1 + {H_1}^2 \right)
 - \frac{ \Lambda^2}{3\beta^2 \kappa^2} \left( 1 - \e^{- \beta \left( \frac{12 {H_0}^2 + 6 H_1 }{2\Lambda}\right)} \right)
\e^{ \beta \left( \frac{12 {H_0}^2 + 6 H_1 }{2\Lambda}\right)} \nonumber \\
&\, - \frac{\Lambda}{\beta\kappa^2} \left( {H_0}^2 + H_1 \right)
 - \frac{\Lambda}{3\beta^2 \kappa^2} \left( 3{H_0}^2 + H_1 \right)\e^{ \beta \left( \frac{12 {H_0}^2 + 6 H_1 }{2\Lambda}\right)} \, .
\end{align}
Then for example, when $\Lambda$ is small enough, $\rho_{{f(R)}\,
(2)} \to \frac{3}{\kappa^2} \left( - {H_0}^2 H_1 + {H_1}^2 \right)
>0$. We should note that $H_1$ is negative due to (\ref{H1}). Then in
the case that $\Lambda$ is small enough, the standard phantom
crossing could occur. On the other hand, when $\Lambda$ is large
enough, we find $\rho_{{f(R)}\, (2)} \to
 - \frac{ \Lambda^2}{3\beta^2 \kappa^2} \left( 1 - \e^{- \beta \left( \frac{12 {H_0}^2 + 6 H_1 }{2\Lambda}\right)} \right) <0$,
and as long as $-H_1< 2{H_0}^2$, the inverse phantom crossing
could occur.

Let us consider more the formulation in Section~\ref{SecIII}. By
using (\ref{Cr4}) and (\ref{rhopDM}), the Friedmann equation in
(\ref{1F}) has the following form,
\begin{align}
\label{fFrR}
0=&\, \frac{1}{2\kappa^2} \left( R + f(R) \right) - m(R) n_\mathrm{DM} \nonumber \\
&\, + \left\{ 3\left(\dot H + H^2\right) + 3H \frac{d}{dt} \right\}
\left\{ \frac{1}{2\kappa^2} \left( 1 + f'(R) \right) - m'(R) n_\mathrm{DM}\right\} \, .
\end{align}
By using the $e$-foldings $N$ defined by $a=\e^N$ by assuming
$a=1$ in the present Universe, we find $n_\mathrm{DM} =
n_\mathrm{DM\, (0)} \e^{-3N}$, where $n_\mathrm{DM\, (0)}$ is the
number density of the dark matter particles in the present
Universe. In terms of the $e$-foldings $N$, Eq.~(\ref{fFrR}) can
be rewritten as,
\begin{align}
\label{fFrR2}
0=&\, \frac{1}{2\kappa^2} \left( R + f(R) \right) - m(R) n_\mathrm{DM\, (0)} \e^{-3N} \nonumber \\
&\, + \left\{ 3\left(H H' + H^2\right) + 3H^2 \frac{d}{dN}
\right\} \left\{ \frac{1}{2\kappa^2} \left( 1 + f'(R) \right) -
m'(R) n_\mathrm{DM\, (0)} \e^{-3N}\right\} \, ,
\end{align}
where $H'\equiv \frac{dH}{dN}$ and we used the relation
$\frac{d}{dt} = H \frac{d}{dN}$. We assume that the time
evolution of the Universe is given by a function $H=H(N)$ and
therefore the scalar curvature $R$ is a function of $N$ because
$R=12 H(N)^2 + 6 H(N) H'(N)$. We may solve $R$ with respect to
$N$, $N=N(R)$. We also assume the $N$-dependence of
$\Phi=\Phi(N)\equiv \frac{1}{2\kappa^2} \left( 1 + f'(R) \right) -
m'(R) n_\mathrm{DM\, (0)} \e^{-3N}$. Then Eq.~(\ref{fFrR2}) gives
\begin{align}
\label{PsiPhi}
\Psi=&\, \Psi(N) \equiv \frac{1}{2\kappa^2} \left( R + f(R) \right) - m(R) n_\mathrm{DM\, (0)} \e^{-3N} \nonumber \\
=&\, - \left\{ 3\left(H H' + H^2\right) \Phi(N) + 3H^2 \frac{d\Phi(N)}{dN} \right\} \, ,
\end{align}
Since,
\begin{align}
\label{PsiPhi2}
\Psi' = \left( 24 H H' + 6 {H'}^2 + 6H H'' \right) \Phi + 3 m(R) n_\mathrm{DM\, (0)} \e^{-3N}\, ,
\end{align}
we find,
\begin{align}
\label{PsiPhi3}
m \left( R\left(N\right) \right) = \frac{\e^{3N}}{3 n_\mathrm{DM\, (0)}} \left( \Psi' - \left( 24 H H' + 6 {H'}^2 + 6H H'' \right) \Phi \right) \, ,
\end{align}
Using the expression of $N=N(R)$, we find $m(R)$ as a function of
$R$. Then by using the definition of $\Psi$, we find,
\begin{align}
\label{PsiPhi4}
f(R) = - R + 2\kappa^2 \Psi\left( N\left( R \right) \right) + 2\kappa^2 m \left( R \right) \e^{-3N(R)}\, .
\end{align}
Then we can find $f(R)$ as a function of $R$. First, we consider a
very trivial example,
\begin{align}
\label{tri1}
H=H_0\, , \quad \Phi=\Phi_0 \, ,
\end{align}
with constants $H_0$ and $\Phi_0$ to check the consistency in the
above formulation. The spacetime is, of course, the de Sitter
spacetime. Then Eqs.~(\ref{PsiPhi}), (\ref{PsiPhi3}), and
(\ref{PsiPhi4}) indicate that,
\begin{align}
\label{tri2}
\Psi= - 3{H_0}^2 \Phi_0\, , \quad m(R)=0 \, , \quad f(R) = - 12 {H_0}^2 - 6 \kappa^2 \Psi_0\, .
\end{align}
Hence we obtained the model with a cosmological constant, where
$f(R)$ is a constant and there is no matter $m(R)=0$, as expected.
We can find the above formulation using Eqs.~(\ref{PsiPhi}),
(\ref{PsiPhi3}), and (\ref{PsiPhi4}) may work. Second, as a less
non-trivial example, we consider the case that,
\begin{align}
\label{tri3}
H^2 = {H_0}^2 + {H_1}^2 \e^{-3N} \, ,
\end{align}
with constants $H_0$ and $H_1$. Then we obtain $R = 12 H^2 + 6 HH'
= 12 \left( {H_0}^2 + {H_1}^2 \e^{-3N} \right) - 9 {H_1}^2
\e^{-3N} = 12{H_0}^2 + 3 {H_1}^2 \e^{-3N}$ and we find
$N=-\frac{1}{3} \ln \left( \frac{R - 12 {H_0}^2}{3{H_1}^2}
\right)$. Again, by using Eqs.~(\ref{PsiPhi}), (\ref{PsiPhi3}),
and (\ref{PsiPhi4}) and assuming $\Phi=\Phi_0$ with a constant
$\Phi_0$, we find,
\begin{align}
\label{tri4}
\Psi=&\, \left( - 3{H_0}^2 + \frac{3}{2} {H_1}^2 \e^{-3N} \right) \Phi_0\, , \quad
m \left( R \right) = \frac{3{H_1}^2 \Phi_0}{2 n_\mathrm{DM\, (0)}}\, , \nonumber \\
f(R) =&\, \left( - 1 + \frac{2\kappa^2 \Phi_0}{n_\mathrm{DM\, (0)}} \right) R + 6\kappa^2 {H_0}^2 \Phi_0 \left( - 1 + \frac{4}{n_\mathrm{DM\, (0)}} \right)\, .
\end{align}
Therefore, we obtain Einstein's gravity with a cosmological
constant consistently although the gravitational coupling $\kappa$
is effectively modified as $\kappa^2 \to \frac{n_\mathrm{DM\,
(0)}}{\Phi_0}$ because $F(R) = R + f(R) = \frac{2\kappa^2
\Phi_0}{n_\mathrm{DM\, (0)}} R + 6\kappa^2 {H_0}^2 \Phi_0 \left( -
1 + \frac{4}{n_\mathrm{DM\, (0)}} \right)$. We now consider a more
realistic and non-trivial model. As we found, the inverse phantom
crossing occurs when the first derivative of the effective DE
density $\rho^\mathrm{eff}_\mathrm{DE}$ vanishes and its second
derivative is negative,
\begin{align}
\label{ipc}
\dot\rho^\mathrm{eff}_\mathrm{DE}=0\, , \quad \ddot\rho^\mathrm{eff}_\mathrm{DE} < 0 \, .
\end{align}
The $e$-foldings $N$ is a monotonically increasing function of the
cosmic time $t$, and the above condition (\ref{ipc}) can be
rewritten in terms of $N$, as follows,
\begin{align}
\label{ipc2}
{\rho^\mathrm{eff}_\mathrm{DE}}'=0\, , \quad {\rho^\mathrm{eff}_\mathrm{DE}}'' < 0 \, .
\end{align}
As an example, we consider the following model,
\begin{align}
\label{ipc3}
\rho^\mathrm{eff}_\mathrm{DE} 
= \Lambda_0 + \frac{\Lambda_1}{ \left( \e^{-3N} - \e^{-3N_0} \right)^2 + C^2} \, .
\end{align}
Here $\Lambda_0$ and $\Lambda_1$ are constants with the dimension
of $[m]^4$, and $C$ is a dimensionless constant. The effective
energy density $\rho^\mathrm{eff}_\mathrm{DE}$ has a peak at
$N=N_0$, which corresponds to the $e$-foldings that the inverse
phantom crossing occurs.

We also assume that the effective dark matter density
$\rho^\mathrm{eff}_\mathrm{DM}$ behaves as
$\rho^\mathrm{eff}_\mathrm{DM} \propto a^{-3}=\e^{-3N}$,
\begin{align}
\label{ipc3B}
\rho^\mathrm{eff}_\mathrm{DM} = \rho_\mathrm{DM\, (0)} \e^{-3N}\, .
\end{align}
Then the Friedmann equation,
\begin{align}
\label{1Fe2}
\frac{3}{\kappa^2} H^2 = \rho^\mathrm{eff}_\mathrm{DE} + \rho^\mathrm{eff}_\mathrm{DM} \, ,
\end{align}
indicates that,
\begin{align}
\label{Ht1}
H^2 = \frac{3}{\kappa^2} \left\{ \Lambda_0 + \frac{\Lambda_1}{ \left( \e^{-3N} - \e^{-3N_0} \right)^2 + C^2} + \rho_\mathrm{DM\, (0)} \e^{-3N} \right\} \, .
\end{align}
Therefore, the scalar curvature $R$ is given by,
\begin{align}
\label{Ht2}
R =&\, 12 H^2 + 6 H H' \nonumber \\
=&\, \frac{36}{\kappa^2} \left\{ \Lambda_0 + \frac{\Lambda_1}{ \left( \e^{-3N} - \e^{-3N_0} \right)^2 + C^2} + \rho_\mathrm{DM\, (0)} \e^{-3N} \right\} \nonumber \\
&\, + \frac{9}{\kappa^2} \left\{ \frac{6\Lambda_1\left( \e^{-3N} - \e^{-3N_0} \right)\e^{-3N}}{ \left\{ \left( \e^{-3N} - \e^{-3N_0} \right)^2 + C^2\right\}^2 }
 - 3 \rho_\mathrm{DM\, (0)} \e^{-3N} \right\} \, .
\end{align}
In principle, Eq.~(\ref{Ht2}) can be solved with respect to $N$,
$N=N(R)$, although it is difficult to do it explicitly. The
phantom crossing occurs when $N\sim N_0$, given by,
\begin{align}
\label{Ht3}
R =&\, R_0 + R_1 \left( N - N_0 \right) + \mathcal{O} \left( \left( N - N_0 \right)^2 \right) \, , \nonumber \\
& R_0 \equiv \frac{9}{\kappa^2} \left( 4\Lambda_0 + \frac{4\Lambda_1}{C^2}
+ \rho_\mathrm{DM\, (0)} \e^{-3N_0} \right) \, , \quad
R_1 \equiv - \frac{27}{\kappa^2} \left( \frac{30\Lambda_1\e^{-3N_0}}{C^4}
+ \rho_\mathrm{DM\, (0)} \right) \, ,
\end{align}
which can be solved with respect to $N$,
\begin{align}
\label{Ht4}
N=N_0 + \frac{R - R_0}{R_1} + \mathcal{O} \left( \left( R - R_0 \right)^2 \right) \, .
\end{align}
In principle, by using Eqs.~(\ref{PsiPhi}), (\ref{PsiPhi3}), and
(\ref{PsiPhi4}) and properly assuming $\Phi=\Phi_0$, we may find
the form of $f(R)$ and $m(R)$ although it is very difficult to
find their explicit forms.

\section{Oscillating Hubble Rate}

As in \cite{Elizalde:2011ds, Elizalde:2010ts}, we consider the
model where the Hubble rate $H$ oscillates and therefore the
Universe iterates the phantom crossing and the inverse phantom
crossing. In this section, we consider such a oscillation for the
total energy density or the behavior of $H$. A simple model of
this sort is given by,
\begin{align}
\label{Hex1} H^2 = {H_0}^2 \left( 1 + C \sin \left( \omega N
\right) \right)\, ,
\end{align}
where $H_0$, $C$, and $\omega$ are constants, and we assume
$0<C<1$ and $\omega>0$. Then we find,
\begin{align}
\label{Hex2}
HH' = \frac{{H_0}^2 C \omega}{2} \cos \left( \omega N \right)
\, .
\end{align}
Because we are considering the expanding Universe, where $H>0$,
when $-\frac{\pi}{2} + 2n\pi < \omega N < \frac{\pi}{2} + 2n\pi$
with an integer $n$, $H'$ is positive and therefore, the Universe
is in the phantom phase. On the other hand, when $\frac{\pi}{2} +
2n\pi < \omega N < \frac{3\pi}{2} + 2n\pi$ with an integer $n$,
$H'$ is negative and therefore, the Universe is in the non-phantom
phase. The phantom crossing occurs when $\omega N=-\frac{\pi}{2} +
2n\pi$ with an integer $n$ and the inverse phantom crossing occurs
when $\omega N=\frac{\pi}{2} + 2n\pi$ with an integer $n$. Under
the assumption (\ref{Hex1}), the scalar curvature $R$ is given by
\begin{align}
\label{OsH1}
R=&\, 12 H^2 + 6 HH'
= H_0^2 \left( 12 + 12 C \sin \left( \omega N \right) + 3 C \omega \cos \left( \omega N \right) \right)
\nonumber \\
=&\, 3{H_0}^2 \left\{ 4 + \sqrt{16 + \omega^2} C \left( \frac{4}{\sqrt{16 + \omega^2}} \sin \left( \omega N \right)
+ \frac{\omega}{\sqrt{16 + \omega^2}} \cos \left( \omega N \right) \right) \right\} \nonumber \\
=&\, 3{H_0}^2 \left( 4 + \sqrt{16 + \omega^2} C \sin \left( \omega N + \theta_0 \right) \right) \, .
\end{align}
Here $\theta_0$ is defined by $\tan \theta_0 = \frac{\omega}{4}$.
Eq.~(\ref{OsH1}) can be solved with respect to $N$,
\begin{align}
\label{OsH2}
N= - \frac{\theta_0}{\omega} \sin^{-1} \left( \frac{ R - 12 {H_0}^2}{3{H_0}^2 C\sqrt{ 16 + \omega^2}} \right)\, .
\end{align}
Here $sin^{-1}$ is the inverse function of $\sin$ function. The
$e$-foldings $N$ is explicitly given as a function of $R$.
Furthermore, we find,
\begin{align}
\label{sc}
\sin\left(\omega N \right) =&\, \sin\left( \omega N + \theta_0 - \theta_0 \right)
= \sin\left( \omega N + \theta_0 \right) \cos\left( \theta_0 \right) - \cos\left( \omega N + \theta_0 \right)\sin\left( \theta_0 \right) \nonumber \\
=&\, \frac{4}{\sqrt{16 + \omega^2}} \left( \frac{ R - 12 {H_0}^2}{3{H_0}^2 C\sqrt{ 16 + \omega^2}} \right)
 - \frac{\omega}{\sqrt{16 + \omega^2}} \sqrt{ 1 - \left( \frac{ R - 12 {H_0}^2}{3{H_0}^2 C\sqrt{ 16 + \omega^2}} \right) ^2} \, , \nonumber \\
\cos\left(\omega N \right) =&\, \cos\left( \omega N + \theta_0 - \theta_0 \right)
= \cos\left( \omega N + \theta_0 \right) \cos\left( \theta_0 \right) + \sin\left( \omega N + \theta_0 \right)\sin\left( \theta_0 \right) \nonumber \\
=&\, \frac{4}{\sqrt{16 + \omega^2}} \sqrt{ 1 - \left( \frac{ R - 12 {H_0}^2}{3{H_0}^2 C\sqrt{ 16 + \omega^2}} \right) ^2}
+ \frac{\omega}{\sqrt{16 + \omega^2}} \left( \frac{ R - 12 {H_0}^2}{3{H_0}^2 C\sqrt{ 16 + \omega^2}} \right) \, ,
\end{align}
which we will use shortly. Again, we use Eqs.~(\ref{PsiPhi}),
(\ref{PsiPhi3}), and (\ref{PsiPhi4}). Then by assuming
$\Phi=\Phi_0$, we find,
\begin{align}
\label{OsH3}
\Psi (N) =&\, - 3 \Phi_0 {H_0}^2 \left( 1 + C \sin \left( \omega N \right) + \frac{C \omega}{2} \cos \left( \omega N \right) \right) \, , \nonumber \\
m(N)
=&\, \frac{\Phi_0 {H_0}^2 \e^{3N}}{3 n_\mathrm{DM\, (0)}} \left\{ - 3 \left( C \omega \cos \left( \omega N \right) - \frac{C \omega^2}{2} \sin \left( \omega N \right) \right)
 - 12 C \omega \cos \left( \omega N \right) + 3 C \omega^2 \sin \left( \omega N \right) \right\} \nonumber \\
=&\, - \frac{\Phi_0 {H_0}^2 C \e^{3N}}{n_\mathrm{DM\, (0)}} \left\{ \frac{9\omega^2}{2} \sin \left( \omega N \right)
 - 15 \omega \cos \left( \omega N \right) \right\} \, , \nonumber \\
f
=&\, - R + 2 \kappa^2 \left\{ - 3 \Phi_0 {H_0}^2 \left( 1 + C \sin \left( \omega N \right) + \frac{C \omega}{2} \cos \left( \omega N \right) \right) \right\} \nonumber \\
&\, - \frac{2\kappa^2 \Phi_0 {H_0}^2 C} {n_\mathrm{DM\, (0)}} \left\{ \frac{9\omega^2}{2} \sin \left( \omega N \right)
 - 15 \omega \cos \left( \omega N \right) \right\} \nonumber \\
=&\, - R - 2\kappa^2 \Phi_0 {H_0}^2 \left[ 3 + C \left\{ \left( 3 + \frac{9 \omega^2}{2 n_\mathrm{DM\, (0)}} \right) \sin \left( \omega N \right) \right. \right. \nonumber \\
&\, \left. \left. + \left( \frac{3}{2} - \frac{15}{n_\mathrm{DM\, (0)}} \right) \omega \cos \left( \omega N \right) \right\} \right] \, .
\end{align}
Then by using (\ref{OsH2}), we find the $R$ dependence of $m=m(R)$
and $f=f(R)$,
\begin{align}
\label{OsH4}
m(R)
=&\, - \frac{\Phi_0 {H_0}^2 C \e^{- \frac{3\theta_0}{\omega} \sin^{-1} \left( \frac{ R - 12 {H_0}^2}{3{H_0}^2 C \sqrt{ 16 + \omega^2}} \right)}}
{n_\mathrm{DM\, (0)}} \left[ \frac{9\omega^2}{2}
\left\{ \frac{4}{\sqrt{16 + \omega^2}} \left( \frac{ R - 12 {H_0}^2}{3{H_0}^2 C\sqrt{ 16 + \omega^2}} \right) \right. \right. \nonumber \\
&\, \left. - \frac{\omega}{\sqrt{16 + \omega^2}} \sqrt{ 1 - \left( \frac{ R - 12 {H_0}^2}{3{H_0}^2 C\sqrt{ 16 + \omega^2}} \right) ^2}
\right\}
 - 15 \omega
\left\{ \frac{4}{\sqrt{16 + \omega^2}} \sqrt{ 1 - \left( \frac{ R - 12 {H_0}^2}{3{H_0}^2 C\sqrt{ 16 + \omega^2}} \right) ^2} \right. \nonumber \\
&\, \left. \left. + \frac{\omega}{\sqrt{16 + \omega^2}} \left( \frac{ R - 12 {H_0}^2}{3{H_0}^2 C\sqrt{ 16 + \omega^2}} \right) \right\} \right]
 \, , \nonumber \\
=&\, - \frac{\Phi_0 {H_0}^2 C \e^{- \frac{3\theta_0}{\omega} \sin^{-1} \left( \frac{ R - 12 {H_0}^2}{3{H_0}^2 C \sqrt{ 16 + \omega^2}} \right)}}
{n_\mathrm{DM\, (0)}\sqrt{16 + \omega^2}}
\left\{ 3\omega^2 \left( \frac{ R - 12 {H_0}^2}{3{H_0}^2 C\sqrt{ 16 + \omega^2}} \right) \right. \nonumber \\
&\, \left. - \left( \frac{9\omega^3}{2} + 60\omega \right) \sqrt{ 1 - \left( \frac{ R - 12 {H_0}^2}{3{H_0}^2 C\sqrt{ 16 + \omega^2}} \right) ^2}
\right\}
\, , \nonumber \\
f (R)
=&\, - R - 2\kappa^2 \Phi_0 {H_0}^2
\left[ 3 + C \left\{ \left( 3 + \frac{9 \omega^2}{2 n_\mathrm{DM\, (0)}} \right)
\left\{ \frac{4}{\sqrt{16 + \omega^2}} \left( \frac{ R - 12 {H_0}^2}{3{H_0}^2 C\sqrt{ 16 + \omega^2}} \right) \right. \right. \right. \nonumber \\
&\, \left. - \frac{\omega}{\sqrt{16 + \omega^2}} \sqrt{ 1 - \left( \frac{ R - 12 {H_0}^2}{3{H_0}^2 C\sqrt{ 16 + \omega^2}} \right) ^2}
\right\} \nonumber \\
&\, + \left( \frac{3}{2} - \frac{15}{n_\mathrm{DM\, (0)}} \right) \omega
\left\{ \frac{4}{\sqrt{16 + \omega^2}} \sqrt{ 1 - \left( \frac{ R - 12 {H_0}^2}{3{H_0}^2 C\sqrt{ 16 + \omega^2}} \right) ^2} \right. \nonumber \\
&\, \left. \left. + \frac{\omega}{\sqrt{16 + \omega^2}} \left( \frac{ R - 12 {H_0}^2}{3{H_0}^2 C\sqrt{ 16 + \omega^2}} \right) \right\} \right] \nonumber \\
=&\, - R - 2\kappa^2 \Phi_0 {H_0}^2
\left[ 3 + \frac{C}{\sqrt{16 + \omega^2}} \left\{ \left( 12 + \frac{3}{2} \omega^2 + \frac{3 \omega^2}{n_\mathrm{DM\, (0)}} \right)
\left( \frac{ R - 12 {H_0}^2}{3{H_0}^2 C\sqrt{ 16 + \omega^2}} \right) \right. \right. \nonumber \\
&\, \left. \left. + \left( 3 \omega - \frac{120 \omega + 9 \omega^3}{2 n_\mathrm{DM\, (0)}} \right)
\sqrt{ 1 - \left( \frac{ R - 12 {H_0}^2}{3{H_0}^2 C\sqrt{ 16 + \omega^2}} \right) ^2} \right\} \right] \nonumber \\
\end{align}
In this model, the mass $m=m(R)$ of the dark matter depends on the scalar curvature $R$ non-trivially.

\section{Scalaron as Dark Matter with Curvature-dependent Mass}

There is a scenario in which the dark matter could be a scalar
mode coming from the $F(R)$ gravity \cite{Nojiri:2008nt,
Katsuragawa:2016yir,Katsuragawa:2017wge}. In this scenario, the
mass of the scalar mode naturally depends on the curvature. Then
the particles corresponding to the scalar mode might become dark
matter, which may explain the DESI observation.

The action of the $F(R)$ gravity is given by replacing the scalar
curvature $R$ in the Einstein-Hilbert action which is,
\begin{align}
\label{JGRG6} S_\mathrm{EH}=\int d^4 x \sqrt{-g} \left(
\frac{R}{2\kappa^2} + \mathcal{L}_\mathrm{matter} \right)\, ,
\end{align}
by using some appropriate function of the scalar curvature, as
follows,
\begin{align}
\label{JGRG7} S_{F(R)}= \int d^4 x \sqrt{-g} \left(
\frac{F(R)}{2\kappa^2} + \mathcal{L}_\mathrm{matter} \right)\, .
\end{align}
In Eqs.~(\ref{JGRG6}) and (\ref{JGRG7}),
$\mathcal{L}_\mathrm{matter}$ is the matter Lagrangian density.

By varying the action (\ref{JGRG7}) with respect to the metric, we
obtain the equation of motion for the $F(R)$ gravity theory as
follows,
\begin{align}
\label{JGRG13} \frac{1}{2}g_{\mu\nu} F(R) - R_{\mu\nu} F'(R) -
g_{\mu\nu} \Box F'(R) + \nabla_\mu \nabla_\nu F'(R) = -
\frac{\kappa^2}{2}T_{\mathrm{matter}\,\mu\nu}\, .
\end{align}
For a spatially flat FLRW Universe, Eq.~(\ref{JGRG13}) gives the
field equations,
\begin{align}
\label{JGRG15} 0 =& -\frac{F(R)}{2} + 3\left(H^2 + \dot H\right)
F'(R) - 18 \left( 4H^2 \dot H + H \ddot H\right) F''(R)
+ \kappa^2 \rho_\mathrm{matter}\, ,\\
\label{Cr4b} 0 =& \frac{F(R)}{2} - \left(\dot H + 3H^2\right)F'(R)
+ 6 \left( 8H^2 \dot H + 4 {\dot H}^2 + 6 H \ddot H + \dddot
H\right) F''(R) \nonumber \\ & + 36\left( 4H\dot H + \ddot
H\right)^2 F'''(R) + \kappa^2 p_\mathrm{matter}\, ,
\end{align}
where the Hubble rate $H$ is equal to $H=\dot a/a$. In terms of
the Hubble rate $H$, the scalar curvature $R$ is equal to $R=12H^2
+ 6\dot H$. We now consider the scalar-tensor representation of
the $F(R)$ gravity. We introduce an auxiliary field $A$ and
rewrite the action (\ref{JGRG7}) of the $F(R)$ gravity in the
following form:
\begin{align}
\label{JGRG21} S=\frac{1}{2\kappa^2}\int d^4 x \sqrt{-g}
\left\{F'(A)\left(R-A\right) + F(A)\right\}\, .
\end{align}
We obtain $A=R$ by the variation of the action with respect to $A$
and by substituting the obtained equation $A=R$ into the action
(\ref{JGRG21}), we find that the action in (\ref{JGRG7}) is
reproduced. If we rescale the metric by a kind of a scale
transformation,
\begin{align}
\label{JGRG22} g_{\mu\nu}\to \e^\sigma g_{\mu\nu}\, ,\quad \sigma
= -\ln F'(A)\, ,
\end{align}
we obtain the action in the Einstein frame,
\begin{align}
\label{JGRG23} S_E =& \frac{1}{2\kappa^2}\int d^4 x \sqrt{-g}
\left(
R - \frac{3}{2}g^{\rho\sigma} \partial_\rho \sigma \partial_\sigma \sigma - V(\sigma)\right) \, ,\nonumber \\
V(\sigma) =& \e^\sigma g\left(\e^{-\sigma}\right) - \e^{2\sigma}
f\left(g\left(\e^{-\sigma}\right)\right) =\frac{A}{F'(A)} -
\frac{F(A)}{F'(A)^2}\, .
\end{align}
Here $g\left(\e^{-\sigma}\right)$ is given by solving the equation
$\sigma = -\ln\left( 1 + f'(A)\right)=- \ln F'(A)$ as
$A=g\left(\e^{-\sigma}\right)$. Due to the scale transformation
(\ref{JGRG22}), a coupling of the scalar field $\sigma$ with usual
matter is introduced. The mass of the scalar field $\sigma$ is
given by,
\begin{align}
\label{JGRG24} {m_\sigma}^2 \equiv \frac{3}{2}\frac{d^2
V(\sigma)}{d\sigma^2} =\frac{3}{2}\left\{\frac{A}{F'(A)} -
\frac{4F(A)}{\left(F'(A)\right)^2} + \frac{1}{F''(A)}\right\}\, .
\end{align}
Because $A=R$, the mass depends on the curvature. In the case of
Einstein's gravity, we find $F''(A)=0$ and the mass diverges, as
expected. The scalar mode decouples when $F''(A)=0$.

The scenario that the $F(R)$ scalaron-chameleon can be dark matter
has been investigated \cite{Nojiri:2008nt, Katsuragawa:2016yir,
Katsuragawa:2017wge}. If the mass increases when the curvature $R$
decreases, the problem of the DESI observation might be solved. As
an example, we may consider the following simple model
\begin{align}
\label{ex1} F(R)=R - \Lambda + \alpha R^2 \, .
\end{align}
Then (\ref{JGRG24}) tells,
\begin{align}
\label{ex2} {m_\sigma}^2 = \frac{3}{2} \left\{ \frac{R}{1 +
2\alpha R} - \frac{4\left( R - \Lambda + \alpha R^2
\right)}{\left( 1 + 2\alpha R \right)^2} + \frac{1}{2\alpha}
\right\}\, .
\end{align}
When $R$ is large ${m_\sigma}^2$ goes to vanish and when $R\to 0$,
we find ${m_\sigma}^2 \to 4\Lambda + \frac{1}{2\alpha}$.
Therefore, the mass increases when $R$ decreases.

We now rewrite Eq.~(\ref{ex2}) as follows,
\begin{align}
\label{ex3} {m_\sigma}^2 =&\, \frac{3}{2} \frac{\left( 4 - 8 + 4
\right)\alpha^2 R^2 + \left( 2 - 8 + 4 \right) \alpha R + 8 \alpha
\Lambda + 1}
{2\alpha \left( 1 + 2\alpha R \right)^2} \nonumber \\
=&\, \frac{3 \left( - 2 \alpha R + 8 \alpha \Lambda + 1 \right)}{4
\alpha \left( 1 + 2\alpha R \right)^2} \, .
\end{align}
Therefore, when $R>4\Lambda + \frac{1}{2\alpha}$, the scalar mode
becomes tachyonic. This is not surprising because it corresponds
to the slow-roll during inflation. The model (\ref{ex1}) could
describe inflation due to the last term, $\alpha R^2$, when $R$ is
large. The second term could play the role of a small cosmological
term for the late-time accelerated expansion, when $R$ is small.
Then we expect $\alpha \sim \left( 10^{13}\, \mathrm{GeV}
\right)^{-2}$ and $\Lambda \sim \left( 10^{-33}\, \mathrm{eV}
\right)^2$. Therefore, we may estimate $m_\sigma$ as $m_\sigma
\sim 10^{13}\, \mathrm{GeV}$, which is large enough. In any case,
the model (\ref{ex1}) simply describes both the inflation and the
DE.

\section{Towards a Realistic Model: A Theoretical Approach}

In order to obtain a realistic model, we need to clarify the
creation of the scalaron particles and to check the consistency of
the parameters. First, we should note that, as is clear from the
scale transformation~(\ref{JGRG22}), there is an interaction
between the scalaron and the matter, including the standard model
particles. Then, in the thermal equilibrium, the scalaron could be
created by the interaction. Because the coupling is given by the
gravitational coupling, which is the inverse of the Planck mass,
the interaction is very weak and therefore a sufficient amount of
the scalaron particles could be created, and the mass of the
scalaron must be small enough. Eq.~(\ref{ex2}) could tell that the
mass might be small just after the inflationary era. There might
also be gravitational particle production.

We should note that the DESI observation corresponds to the
redshift region $0<z<3.5$. If we choose the parameter $\alpha$ in
the model in (\ref{ex1}) to be $\alpha \sim \left( 10^{13}\,
\mathrm{GeV} \right)^{-2}$ so that the model could also describe
the inflationary era, Eq.~(\ref{ex3}) tells that in the DESI
redshift, the mass is already heavy enough, ${m_\sigma}^2\sim
\frac{3}{4\alpha}$, and therefore the inverse phantom crossing
could not occur. In order to avoid this problem, we may choose
$1/\alpha$ to be small enough. In the laboratory, the energy
density of the redshift region $0<z<3.5$ can be easily realized,
and the fifth force experiment excludes such a light mode. The
coupling between the scalaron and the matter shifts the mass due
to the Chameleon mechanism, but the scalaron could be light enough
to be excluded from the experiments. In order to avoid this
problem, we may modify the model so that $\alpha$ could depend on
the scalar curvature. The problem might be solved if $1/\alpha$ is
large enough even in the redshift region $0<z<3.5$ to be
consistent with the fifth force experiment and after that
$1/\alpha$ further increases to generate the apparent phantom
crossing.

\section{Some Explicit Examples of Oscillating DE with Phantom Crossing}

\begin{figure}
\centering
\includegraphics[width=18pc]{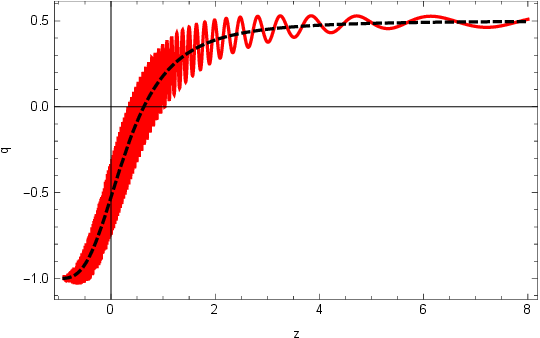}
\includegraphics[width=18pc]{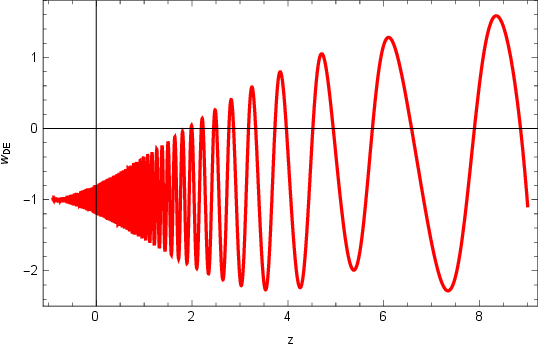}
\includegraphics[width=18pc]{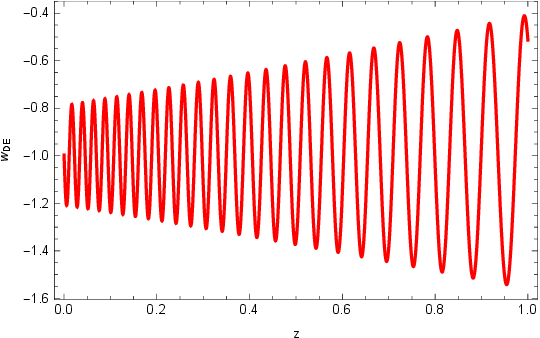}
\caption{Plots of the deceleration parameter $q(z)$ (upper left
plot) and the DE EoS parameter $w_\mathrm{DE}(z)$ (right and bottom
plot) as functions of the redshift for the model Eq. (\ref{fr24})
for $b=20\Lambda$, $c=2$, and $R_0={m_s}^2/0.00091$.}\label{24pl}
\end{figure}

\begin{figure}
\centering
\includegraphics[width=18pc]{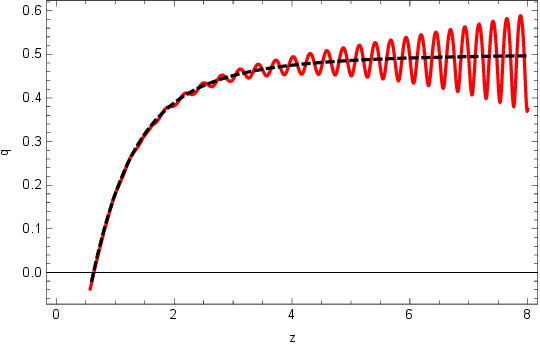}
\includegraphics[width=18pc]{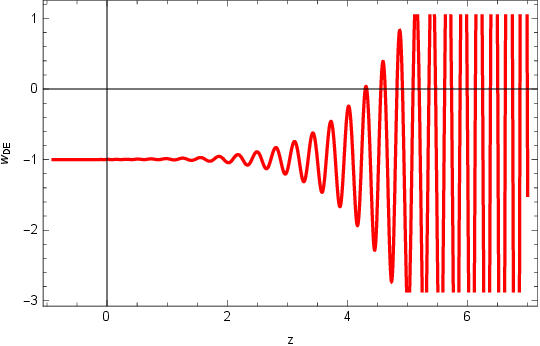}
\includegraphics[width=18pc]{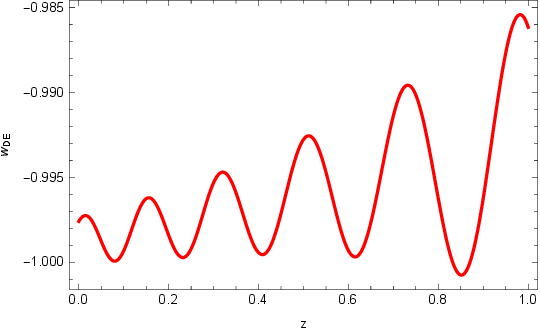}
\caption{Plots of the deceleration parameter $q(z)$ (upper left
plot) and the DE EoS parameter $w_\mathrm{DE}(z)$ (right and bottom
plot) as functions of the redshift for the model of Eq.
(\ref{frhu}) with $\alpha=1.4 \Lambda$, $b=1$, $c=0.2$ , $d=0.04$,
$R_0={m_s}^2$, and $n=0.3$.}\label{hupl}
\end{figure}

Now let us consider some realistic models of $F(R)$ gravity DE
which we will extensively study using numerical methods and we
will obtain the cosmological parameters of these models. We will
be mainly interested in the DE density and DE EoS parameter which
are constrained by the data and we will numerically show that the
models exhibit late-time EoS oscillations and a phantom crossing.
Let us consider the $F(R)$ gravity action in the presence of
perfect matter fluids in (\ref{JGRG7}). Again we assume that the
$F(R)$ gravity has the form of (\ref{FRf}) and upon varying the
action (\ref{JGRG7}) with respect to the FLRW metric, we obtain,
\begin{align} \label{Friedman}
 3 F_R H^2=\kappa^2 \rho_\mathrm{matter} + \frac{F_R R - F}{2} -3H \dot F_R \, ,
\end{align}
\begin{align} \label{Raycha}
 -2 F_R \dot H = \kappa^2 (\rho_\mathrm{matter} + p_\mathrm{matter}) + \ddot F -H \dot F \, ,
\end{align}
with $F_R = \frac{\partial F}{\partial R}$ and the ``dot''
indicates differentiation with respect to cosmic time, and
$\rho_\mathrm{matter}$ and $p_\mathrm{matter}$ stand for the matter fluids energy density and
pressure. We can rewrite the field equations
(\ref{Friedman}),(\ref{Raycha}) in the Einstein-Hilbert gravity
form,
\begin{align} \label{Friedtot}
 3 H^2= \kappa^2 \rho_\mathrm{total} \ ,
\end{align}
\begin{align} \label{Raychtot}
 -2 \dot H =\kappa^2 (\rho_\mathrm{total} + p_\mathrm{total}) \ ,
\end{align}
where $\rho_\mathrm{total}$ stands for the total energy density of the
total effective cosmological fluid, which includes the $F(R)$
gravity contribution, and $p_\mathrm{total}$ denotes the total pressure.
The total cosmological fluid is composed by three parts, the cold
dark matter one ($\rho_\mathrm{matter}$), the radiation ($\rho_\mathrm{r}$) and the
geometric part of $F(R)$ gravity ($\rho_\mathrm{DE}$). Therefore, we
have, $\rho_\mathrm{total}=\rho_\mathrm{matter} + \rho_\mathrm{r} + \rho_\mathrm{DE}$ and in addition
$p_\mathrm{total}=p_\mathrm{matter} + p_\mathrm{r} + p_\mathrm{DE}$. It is the geometric fluid which
drives the dynamical evolution in the late-time era, and with its
energy density and pressure being,
\begin{align}\label{rDE}
 \rho_\mathrm{DE}=\frac{F_R R - F}{2} + 3 H^2 (1-F_R)-3H \dot F_R \, ,
\end{align}
\begin{align}\label{PDE}
 p_\mathrm{DE}=\ddot F -H \dot F +2 \dot H (F_R -1) - \rho_\mathrm{DE} \, .
\end{align}
In the following, instead of the cosmic time, we shall use the
redshift as a dynamical variable, defined as,
\begin{align}
 1+z=\frac{1}{a}\, ,
\end{align}
and in addition we introduce the statefinder parameter $y_H (z)$
\cite{Hu:2007nk,Bamba:2012qi,reviews1},
\begin{align} \label{yhdef}
 y_H(z)=\frac{\rho_\mathrm{DE}}{\rho_\mathrm{matter}^{(0)}}=\frac{H^2}{{m_s}^2}-(1+z)^3-\chi (1+z)^4 ,
\end{align}
with $\rho_\mathrm{matter}^{(0)}$ denoting the energy density of the cold dark
matter at the present epoch, and in addition ${m_s}^2=\frac{\kappa^2
\rho_\mathrm{matter}^{(0)}}{3}=H_0^2 \Omega_\mathrm{matter}=1.37 \times 10^{-67}\, \mathrm{eV}^2$ while
$\chi$ is $\chi=\frac{\rho_\mathrm{r}^{(0)}}{\rho_\mathrm{matter}^{(0)}} \simeq 3.1
\times 10^{-4}$, with $\rho_\mathrm{r}^{(0)}$ denoting the present epoch
radiation energy density. Combining Eqs.~(\ref{FRf}), (\ref{Friedtot}), and (\ref{yhdef}), we can rewrite the Friedmann
equation in terms of the statefinder $y_H$ in the following way,
\begin{align} \label{difyh}
 \frac{d^2 y_H}{dz^2} + J_1 \frac{d y_H}{dz} + J_2 y_H + J_3=0 \, ,
\end{align}
where the dimensionless functions $J_1$ , $J_2$ , $J_3 $ are
defined as,
\begin{align} \label{J1}
 J_1 = \frac{1}{(z+1)} \Big( -3-\frac{1}{y_H + (z+1)^3 + \chi (z+1)^4} \frac{1-F_R}{6 {m_s}^2F_{RR}}\Big) \, ,
\end{align}

\begin{align} \label{J2}
 J_2 = \frac{1}{(z+1)^2} \Big( \frac{1}{y_H + (z+1)^3 +\chi (z+1)^4} \frac{2-F_R}{3 {m_s}^2 F_{RR}}\Big) \, ,
\end{align}

\begin{align} \label{J3}
 J_3 = -3(z+1) - \frac{(1-F_R)((z+1)^3 + 2\chi (z+1)^4) + (R-F)/(3 {m_s}^2)}{(z+1)^2 (y_H + (z+1)^3 +\chi (z+1)^4)} \frac{1}{6 {m_s}^2 F_{RR}}\, ,
\end{align}
and furthermore $F_{RR}=\frac{\partial^2 F}{\partial R^2}$.
Moreover, the Ricci scalar is,
\begin{align}\label{ft10newadd}
R=12H^2-6HH_{z}(1+z)\, ,
\end{align}
and in terms of $y_H$ we obtain,
\begin{align}\label{neweqnrefricciyH}
R(z)=3\,{m_s}^2\left(-(z+1)\,\frac{d y_H(z)}{dz} + 4 y_H(z) +
(1+z)^3\right)\, .
\end{align}
We will solve Eq. (\ref{difyh}) numerically for the late-time
redshift interval $z=[0,10]$, using the following initial
conditions at the final redshift $z_f=10$ \cite{Bamba:2012qi},
\begin{align}\label{initialcond}
 y_H (z_f) = \frac{ \Lambda}{3 {m_s}^2} \Big( 1 + \frac{1+z_f}{1000} \Big) \ , \ \frac{d y_H(z)}{dz} \Big |_{z=z_f} = \frac{1}{1000} \frac{ \Lambda}{3 {m_s}^2},
\end{align}
where $\Lambda \simeq 11.895 \times 10^{-67}\, \mathrm{eV}^2$. We can express
the physical cosmological quantities using the statefinder
$y_H(z)$, and we have,
\begin{align}\label{hubblefr}
H(z)=m_s\sqrt{y_H(z)+(1+z)^{3}+\chi (1+z)^4}\, .
\end{align}
and the Ricci scalar is,
\begin{align}\label{curvature}
 R(z)=3 {m_s}^2 \Big( 4 y_H(z) -(z+1) \frac{d y_H (z)}{dz} + (z+1)^3 \Big),
\end{align}
while the DE density parameter $\Omega_\mathrm{DE}(z)$ is,
\begin{align}\label{OmegaDE}
 \Omega_\mathrm{DE}(z)=\frac{y_H(z)}{y_H(z)+(z+1)^3 + \chi (z+1)^4}\, ,
\end{align}
and the DE EoS parameter is,
\begin{align}
\label{EoSDE}
w_\mathrm{DE}(z)=-1+\frac{1}{3}(z+1)\frac{1}{y_H(z)}\frac{d y_H(z)}{dz}\,
 .
\end{align}
Finally, the total EoS parameter is,
\begin{align}\label{EoStot}
 w_\mathrm{total}(z)=\frac{2(z+1)H'(z)}{3H(z)}-1 \, .
\end{align}
while the deceleration parameter is,
\begin{align}\label{declpar}
 q(z)=-1-\frac{\dot H}{H^2}=-1-(z+1)\frac{H'(z)}{H(z)},
\end{align}
where the ``prime'' denotes differentiation with respect to the
redshift. In addition, and for comparison reasons, let us quote
here the Hubble rate for the $\Lambda$CDM model,
\begin{align}\label{hubblelcdm}
 H_{\Lambda}(z)=H_0\sqrt{\Omega_{\Lambda} +\Omega_\mathrm{matter}(z+1)^3 +\Omega_\mathrm{r}(z+1)^4 } ,
\end{align}
where $\Omega_{\Lambda} \simeq 0.68136$ and
$\Omega_\mathrm{matter} \simeq 0.3153$ and $ H_0 \simeq 1.37187
\times 10^{-33}\, \mathrm{eV}$ according to the latest 2018 Planck
data \cite{Planck:2018vyg}. Now let us focus on the phenomenology
of two interesting models which exhibit remarkable behavior at
late times. We start off with the following model,
\begin{align}\label{fr24}
 F(R)=R+\frac{R^2}{M^2}-\frac{b}{c+\exp(-R/R_0)},
\end{align}
with $b$ and $R_0$ are free parameters with mass dimensions
$[m]^2$ and also $c$ is a dimensionless parameter. We obtained a
viable phenomenology for the choice $b=20\Lambda,$ $c=2$, and
$R_0={m_s}^2/0.00091$. In Fig. \ref{24pl} we plot the deceleration
parameter for this model (red curve) compared with the
$\Lambda$CDM (black curve), and the DE EoS for various redshift
ranges. As it can be seen, the DE EoS is strongly oscillating at
late times and the model ends up in a quintessential regime. The
DE energy density parameter and the DE EoS parameter at $z=0$ are
for this model $\Omega_\mathrm{DE}=0.6918$ and $w_\mathrm{DE}=-0.9974$
which comply with the latest Planck constraints.
Another phenomenologically interesting $F(R)$ model is,
\begin{align}\label{frhu}
 F(R)=R+\frac{R^2}{M^2}-\alpha \frac{b(R/R_0)^n}{c(R/R_0)^n+d},
\end{align}
with $b,c,d,n$ being dimensionless parameters and $\alpha,R_0$ are
also free parameters with mass dimensions $[m]^2$. With the choice
$\alpha=1.4 \Lambda$, $b=1$, $c=0.2$ , $d=0.04$, $R_0={m_s}^2$,
and $n=0.3$ we obtain an interesting DE oscillating behavior which
is also phenomenologically viable. Specifically, in Fig.
\ref{hupl} we plot the deceleration parameter (red curve) compared
with the $\Lambda$CDM (black curve), and also the DE EoS for
various redshift ranges, focusing at small redshifts. As it can be
seen in this case too, the DE EoS is strongly oscillating at late
times and at small redshifts before present day, and the model
ends up in a quintessential regime in this case too. The DE energy
density parameter and the DE EoS parameter at $z=0$ for this model
are $\Omega_\mathrm{DE}=0.6851$ and $w_\mathrm{DE}=-0.9887$ which
are again compatible with the latest Planck constraints.
Thus in this section using a numerical approach, we demonstrated
that it is possible to obtain oscillating DE eras using rather
simple $F(R)$ gravity models. Note that the models presented in
this section were also extensively examined regarding their
phenomenology in Ref. \cite{Odintsov:2025jfq} and the model
(\ref{fr24}) is found to be fully compatible with the
observational data, including the latest DESI data.

\section{Conclusions}

In this paper we studied the possibility of having dynamical DE
which crosses the phantom divide line and oscillates at late
times. We examined this problem both theoretically and numerically
and proved that $F(R)$ gravity provides a consistent
phenomenological framework for naturally describing the DE era
without resorting to phantom fields. Specifically, we discussed
the inverse-phantom crossing in $F(R)$ gravity from a theoretical
viewpoint and discussed the conditions that the $F(R)$ gravity
fluid must satisfy in order to achieve the inverse phantom
crossing. We also discussed the apparent phantom crossing issue,
further extending the considerations made. Furthermore, we
developed the theoretical study of several models that may exhibit
phantom crossing and discussed from a theoretical viewpoint these
models and their DE behavior, in a qualitative way though.
Moreover, several oscillating forms of the Hubble rate were
considered, in terms of the $e$-foldings number and their
realization in $F(R)$ gravity was also considered. In addition, we
studied the effects of a scalaron as dark matter with a
curvature-dependent mass on the DE era. We also studied in a
quantitative way the predictions on the DE era of two explicit
$F(R)$ gravity models, one of which is also compatible with the
DESI data apart from the Planck data. We considered the behavior
of the deceleration parameter and the DE EoS and showed that it is
possible to obtain a phantom crossing EoS and in addition an
oscillating DE EoS at late times. Thus we proved that indeed
$F(R)$ gravity offers a consistent framework which can generate a
phantom to quintessence transition in a natural way without
phantom fields and moreover some of the models realize DE
oscillations, again built in the model, not artificially added.
Now the question is why should $F(R)$ gravity should be considered
a viable extension of Einstein's gravity. The answer is simple,
Einstein's gravitational Lagrangian density is described simply by
$R$, so an Occam's razor approach is that an $F(R)$ gravity
extends it. Of course, someone may also seek if such oscillating
and phantom divide crossing phenomena occur in other gravities,
like Einstein-Gauss-Bonnet theories, but the natural extension of
Einstein's gravity is $F(R)$ gravity and it enjoys an elevated
role among all modified gravity models.

\section*{Acknowledgements}

This work was partially supported by the program Unidad de
Excelencia Maria de Maeztu CEX2020-001058-M, Spain (S.D.O).

\end{document}